# Lindbladian operators, von Neumann entropy and energy conservation in time-dependent quantum open systems

Congjie Ou [a], Ralph V. Chamberlin [b], Sumiyoshi Abe [a,c,d]

[a] *College of Information Science and Engineering, Huaqiao University, Xiamen 361021, China*
[b] *Department of Physics, Arizona State University, Tempe, AZ 85287-1504, USA*
[c] *Department of Physical Engineering, Mie University, Mie 514-8507, Japan*
[d] *Institute of Physics, Kazan Federal University, Kazan 420008, Russia*

ABSTRACT

The Lindblad equation is widely employed in studies of Markovian quantum open systems. Here, the following question is posed: in a quantum open system with a time-dependent Hamiltonian such as a subsystem in contact with the heat bath, what is the corresponding Lindblad equation for the quantum state that keeps the internal energy of the subsystem constant in time? This issue is of importance in realizing quasi-stationary states of open systems such as quantum circuits and batteries. As an illustrative example, the time-dependent harmonic oscillator is analyzed. It is shown that the Lindbladian operator is uniquely determined with the help of a Lie-algebraic structure, and the time derivative of the von Neumann entropy is shown to be nonnegative if the curvature of the harmonic potential monotonically decreases in time.

1. **Introduction**

Quantum open systems have long been attracting particular attention in connection with a variety of problems such as errors in quantum computation, measurements, decoherence for micro-macro transition, and foundations of statistical mechanics. Accordingly, a lot of effort has been devoted to the study of nonunitary quantum subdynamics. The standard approach is to consider an isolated multipartite system governed by unitary dynamics with given interactions, to identify an objective subsystem, and then to eliminate the remaining environmental degrees of freedom to obtain the subdynamics of the objective subsystem. In cases where interaction and entanglement between the objective subsystem and its environment are not strong, it may be possible to identify a partial Hamiltonian of the subsystem, which however cannot fully describe the subdynamics because of the nonunitarity. A question of interest here is: how is it possible for the internal energy of such an open system to be conserved? This is relevant to characterizing quasi-stationary quantum open systems that are of contemporary importance. An example is found in quantum thermodynamics, where a constant-internal-energy process (refereed to as an isoenergetic process) is different from an isothermal process, in general, because of the quantum-mechanical violation of the law of equipartition of energy that may lead to some exotic properties of quantum heat engines [1]. A couple of other examples are quantum circuits [2] and batteries (see [3] and the references cited therein). The question can drastically be simplified in the Markovian approximation, where the equation becomes that of the Lindblad type [4-7]. And, another issue concerning the quantum subdynamics is the time evolution of the von Neumann entropy.

The purpose of this paper is to study the problems mentioned above. In Sec. 2, conservation of the internal energy is discussed for a quantum open system with a time-dependent Hamiltonian. There, the equation is derived for the Lindbladian operators that conserve the internal energy. In Sec. 3, the example of the time-dependent harmonic oscillator is analyzed in detail. It is shown that the condition of conservation of the internal energy uniquely determines the Lindbladian operators and the time derivative of the von Neumann entropy is nonnegative in accordance with complete positivity of the subdynamics. Section 4 is devoted to concluding remarks. In addition, a succinct review is given in Appendix about time evolution of the von Neumann entropy under the Lindblad equation.

## 2. Lindblad equation and conservation of internal energy of time-dependent system

Consider a quantum open system with a time-dependent Hamiltonian, $H(t)$. The density matrix, $\rho(t)$, describing its state is assumed to obey the Lindblad equation:

$$i\hbar \partial \rho(t)/\partial t = [H(t), \rho(t)] - i \sum_{m,n} a_{mn} \left( Q_m^\dagger Q_n \rho(t) + \rho(t) Q_m^\dagger Q_n - 2 Q_n \rho(t) Q_m^\dagger \right),$$

where $Q_n$'s and $a_{mn}$ are the operators responsible for nonunitarity of the subdynamics and the element of a $c$-number Hermitian matrix, respectively, and in general both of them may also depend explicitly on time. Making use of the $c$-number unitary matrix with the elements $u_{ml}$, we express the $c$-number matrix as $a_{mn} = \sum_l u_{ml} \alpha_l u_{nl}^*$, where $\alpha_l$ is real. Accordingly, defining, $L_l = \sum_m u_{ml} Q_m$ we rewrite the equation as follows:

$$i\hbar \frac{\partial \rho(t)}{\partial t} = [H(t), \rho(t)] - i \sum_n \alpha_n \left( L_n^\dagger L_n \rho(t) + \rho(t) L_n^\dagger L_n - 2 L_n \rho(t) L_n^\dagger \right). \qquad (1)$$

This linear equation preserves the normalization condition, $\mathrm{tr}\, \rho(t) = 1$, and is known to generate a completely positive dynamical semi-group if $\alpha_n$'s are nonnegative [4-7]. The nonnegativity condition is required in order to incorporate any possible quantum state.

The internal energy is given by $E = \langle H(t) \rangle \equiv \mathrm{tr}(H(t) \rho(t))$. We are interested in a physical situation where $E$ is conserved in time. Therefore, we have

$$\mathrm{tr}\left[ \hbar \frac{\partial H(t)}{\partial t} \rho(t) - H(t) \sum_n \alpha_n \left( L_n^\dagger L_n \rho(t) + \rho(t) L_n^\dagger L_n - 2 L_n \rho(t) L_n^\dagger \right) \right] = 0, \qquad (2)$$

where Eq. (1) and $\mathrm{tr}(H(t)[H(t), \rho(t)]) = 0$ have been used. This condition is satisfied if the following equation holds:

$$\hbar \frac{\partial H(t)}{\partial t} = \sum_n \alpha_n \left( L_n^\dagger L_n H(t) + H(t) L_n^\dagger L_n - 2 L_n^\dagger H(t) L_n \right). \qquad (3)$$

Thus, given $H(t)$, we require $L_n$'s to be determined by this general condition.

Mathematically, Eq. (3) may also contain a term that has a vanishing expectation value. An example is $A - \langle A \rangle$, where $A$ is a certain observable. However, we do not consider such a case here since we wish the Hamiltonian to be independent of the quantum state.

From Eq. (3), it is clear that at least one of $L_n$'s should not commute with the Hamiltonian. Also, in the case when the Hamiltonian and $L_n$'s belong to the trace class, it follows from Eq. (3) that the sum of the eigenvalues of the Hamiltonian is constant in time if $\left[L_n^\dagger, L_n\right] = 0$ (i.e., $L_n$'s are *normal*).

It is emphasized that Eq. (3), which may determine $L_n$'s, does not depend on the details of the interaction between the objective subsystem and its environment, implying how conservation of the internal energy sets a stringent condition on the master equation.

## 3. An example: Time-dependent harmonic oscillator

Let us apply the general discussion developed above to a simple but illustrative example of the harmonic oscillator with unit mass and time-dependent spring coefficient, $k(t)$. The Hamiltonian is $H(t) = p^2/2 + k(t)x^2/2$. It is convenient to introduce the operators: $K_1 = p^2/2$, $K_2 = x^2/2$, $K_3 = (xp+px)/2$. These satisfy the commutation relations: $\left[K_1, K_2\right] = -i\hbar K_3$, $\left[K_2, K_3\right] = 2i\hbar K_2$, $\left[K_3, K_1\right] = 2i\hbar K_1$, which show that $K_a$'s form a Lie algebra isomorphic to $su(1,1)$. The Hamiltonian is then expressed as

$$H(t) = K_1 + k(t)K_2. \tag{4}$$

As seen below, $L_n$'s can be chosen to be Hermitian and, accordingly, Eq. (3) is rewritten as follows:

$$\hbar \frac{\partial H(t)}{\partial t} = \sum_n \alpha_n \left[ L_n, \left[ L_n, H(t) \right] \right]. \tag{5}$$

In addition, it turns out to be sufficient to consider only one operator, say $L_1 \equiv L$. That is, $\alpha_1 \equiv \alpha(t)$ $(\alpha_2 = \alpha_3 = \cdots = 0)$. Because of the Lie-algebraic structure, $L$ has the form:

$$L = c_1 K_1 + c_2 K_2 + c_3 K_3, \tag{6}$$

where $c_a$'s are real $c$-numbers. Then, Eq. (5) is calculated to be

$$\dot{k}(t)K_2 = -2\hbar\alpha(t)\left\{ \left(k(t)c_1^2 - c_1c_2 + 2c_3^2\right)K_1 \right.$$
$$\left. -\left(k(t)c_1c_2 - c_2^2 - 2k(t)c_3^2\right)K_2 + \left(k(t)c_1 + c_2\right)c_3 K_3 \right\}, \tag{7}$$

where $\dot{k}(t) \equiv dk(t)/dt$. Therefore, we have the following coupled equations:

$$k(t)c_1^2 - c_1c_2 + 2c_3^2 = 0, \quad \dot{k}(t) = 2\hbar\alpha(t)\left(k(t)c_1c_2 - c_2^2 - 2k(t)c_3^2\right), \quad \left(k(t)c_1 + c_2\right)c_3 = 0.$$

The only nontrivial solution of these equations is: $c_1 = c_3 = 0$, $c_2^2 = -\dot{k}(t)/[2\hbar\alpha(t)]$. Since $c_2$ can always be absorbed in the definition of $\alpha(t)$, we may set it equal to unity. Therefore, we have $L = K_2 = x^2/2$, $\alpha(t) = -\dot{k}(t)/(2\hbar)$. Consequently, we obtain the following master equation:

$$i\hbar \frac{\partial \rho(t)}{\partial t} = [H(t), \rho(t)] + i\frac{\dot{k}(t)}{8\hbar}\left[x^2, [x^2, \rho(t)]\right]. \qquad (8)$$

In order for this dynamics to be completely positive, the condition

$$\dot{k}(t) \leq 0 \qquad (9)$$

should hold.

Equation (8) should be compared with the one discussed in Ref. [8] for formulating quantum dynamics of macroscopic objects. There, the operator corresponding to $L$ is linear with respect to the position operator, whereas $L$ in Eq. (8) is quadratic.

Finally, let discuss how the von Neumann entropy

$$S[\rho] = -\operatorname{tr}(\rho \ln \rho) \qquad (10)$$

evolves in time under the master equation in Eq. (8). Equation (A.1) in Appendix can be written in terms of $\Gamma_1 \equiv \Gamma = \operatorname{tr}\{(\ln \rho)(L^2 \rho - L\rho L)\}$ as

$$\frac{dS}{dt} = -\frac{\dot{k}(t)}{\hbar^2}\Gamma. \qquad (11)$$

Since $L$ is Hermitian, it immediately follows from Eq. (A.3) in Appendix that

$$\Gamma \geq 0. \qquad (12)$$

Therefore, with Eq. (9), we find that $dS/dt \geq 0$. This result may have a simple physical interpretation. Equation (9) implies that the harmonic potential is widening (i.e., expansion), and accordingly the energy spectrum is being lowered. To conserve the internal energy, the oscillator has to absorb the energy from the environment, e.g., heat energy, if the environment is the heat bath.

## 4. Concluding remarks

We have discussed quantum open systems, whose Hamiltonians are dependent on time explicitly but with internal energies being conserved. We have derived the condition on the Lindbladian operators that must be satisfied in such a situation. We have analyzed in detail the time-dependent Harmonic oscillator as an illustrative example and have shown how the corresponding Lindbladian operator can be obtained from the condition of conservation of the internal energy. We have also shown that the time derivative of the von Neumann entropy is nonnegative in accordance with the time dependence of the oscillator Hamiltonian.

**Acknowledgments**

The work of CO was supported by the grants from Fujian Province (No. 2015J01016, No. JA12001, No. 2014FJ-NCET-ZR04) and from Huaqiao University (No.

ZQN-PY114). He also thanks Toka-Donghua Educational and Cultural Exchange Foundation for providing him with the scholarship and Mie University for the hospitality extended to him. SA would like to thank the High-End Foreign Expert Program of China for support and the warm hospitality of Huaqiao University. His work was also supported in part by a Grant-in-Aid for Scientific Research from the Japan Society for the Promotion of Science (No. 26400391) and by the Program of Competitive Growth of Kazan Federal University from the Ministry of Education and Science of the Russian Federation.

*Note added*. Recently, the concept of *weak invariants* has been proposed and studied for time-dependent quantum dissipative systems [9]. From this viewpoint, the Hamiltonian in Eq. (4) can be regarded as a weak invariant associated with the Lindblad equation (8).

**Appendix**

Here, let us discuss how the von Neumann entropy in Eq. (10) evolves in time under the Lindblad equation. It turned out that this issue has already been studied almost 30 years ago [10]. However, it seems convenient for the reader to present a succinct review of the discussion.

Using the normalization condition on the density matrix and Eq. (1) as well as the basic properties of the trace operation, we find

$$\frac{dS}{dt} = \frac{2}{\hbar} \sum_n \alpha_n \Gamma_n, \qquad (A.1)$$

where $\Gamma_n$ is given by

$$\Gamma_n = \text{tr}\left\{ [\ln \rho(t)] \left[ L_n^\dagger L_n \rho(t) - L_n \rho(t) L_n^\dagger \right] \right\}. \qquad (A.2)$$

The purpose here is to show that the quantity in Eq. (A.2) satisfies the following inequality:

$$\Gamma_n \geq \left\langle \left[ L_n^\dagger, L_n \right] \right\rangle \equiv \text{tr}\left\{ \left[ L_n^\dagger, L_n \right] \rho(t) \right\}. \qquad (A.3)$$

To do so, first let us perform the instantaneous diagonalization of the density matrix at time $t$: $\rho(t) = \sum_i p_i(t) |u_i(t)\rangle \langle u_i(t)|$, where $\{|u_i(t)\rangle\}_i$ is a certain complete orthonormal system satisfying $\langle u_i(t) | u_j(t) \rangle = \delta_{ij}$ and $\sum_i |u_i(t)\rangle \langle u_i(t)| = I$ with $I$ being the identity matrix. It is assumed here that the density matrix is positive definite, i.e., $p_i(t) \in (0,1)$ [that should satisfy the normalization condition: $\sum_i p_i(t) = 1$]. Substituting the diagonalized form of the density matrix into Eq. (A.2), we have

$$\Gamma_n = \sum_i (p_i \ln p_i) \langle u_i | L_n^\dagger L_n | u_i \rangle + \sum_{i,j} (-p_j \ln p_i) |\langle u_i | L_n | u_j \rangle|^2. \qquad (A.4)$$

Now, from the inequality $\ln A \leq A - 1$ ($A > 0$) and identification $A = p_i / p_j$, it follows that $-p_j \ln p_i \geq -p_j \ln p_j + p_j - p_i$. Therefore, we find that

$$\Gamma_n \geq \sum_i (p_i \ln p_i) \langle u_i | L_n^\dagger L_n | u_i \rangle + \sum_{i,j} (-p_j \ln p_j + p_j - p_i) |\langle u_i | L_n | u_j \rangle|^2$$

$$= \sum_i (p_i \ln p_i) \langle u_i | L_n^\dagger L_n | u_i \rangle - \sum_{i,j} (p_j \ln p_j) \langle u_j | L_n^\dagger | u_i \rangle \langle u_i | L_n | u_j \rangle$$

$$+ \sum_{i,j} (p_j - p_i) \langle u_j | L_n^\dagger | u_i \rangle \langle u_i | L_n | u_j \rangle$$

$$= \sum_i p_i \langle u_i | [L_n^\dagger, L_n] | u_i \rangle, \tag{A.5}$$

which proves Eq. (A.3).

It should be noted that the entropy rate given in Eq. (A1) is the sum of $\Gamma_n$'s weighted by $\alpha_n$'s and therefore its sign is not immediately determined by the sign of each $\Gamma_n$.

In a recent work [11], Eqs. (A.1) and (A.3) have been generalized for the Rényi entropy. Therefore, now they are seen to be the simple limiting cases of the results given there.